\documentclass[12pt]{article}
\usepackage[mathscr]{eucal}
\usepackage{amssymb,latexsym}
\usepackage{graphicx}

\newtheorem{thm}{Theorem}[section]

\newtheorem{prop}[thm]{Proposition}
\newtheorem{cor}[thm]{Corollary}
\newcommand{\qed}{\hfill $\Box$ \medskip}
\newcommand{\proof}{\noindent \emph{Proof:\ }}

\newcommand\calL{{\mathcal{L}}}
\newcommand\calO{{\mathcal{O}}}

\newcommand\calM{{\mathcal{M}}}
\newcommand\calT{{\mathcal{T}}}
\newcommand\calI{{\mathcal{I}}}

\newcommand\tl{\Theta_{(\ell)}}

\newcommand\calN{{\mathcal{N}}}

\renewcommand\S{\Sigma}

\renewcommand\d{\partial}

\newcommand\D{\nabla}
\newcommand\e{\epsilon}

\newcommand\g{\gamma}

\renewcommand\th{\theta}
\newcommand\scri{\mathscr{I}}

\newcommand\beq{\begin{eqnarray}}
\newcommand\eeq{\end{eqnarray}}
\newcommand\ben{\begin{enumerate}}
\newcommand\een{\end{enumerate}}
\newcommand\bit{\begin{itemize}}
\newcommand\eit{\end{itemize}}

\newcounter{mnotecount}[section]

\title{Some Uniqueness Results for Dynamical Horizons }

\author{
Abhay Ashtekar \thanks{email: ashtekar@gravity.psu.edu}  \\
Institute for Gravitational Physics and Geometry \\ Physics
Department, Penn State, University Park, PA 16802-6300  \\ \\
Gregory J. Galloway\thanks{email: galloway@math.miami.edu} \\
Department of Mathematics \\ University of Miami, Coral Gables, FL
33124}

\begin{document}
\date{}
\maketitle

\begin{abstract}

We first show that the intrinsic, geometrical structure of a
dynamical horizon is unique. A number of physically interesting
constraints are then established on the location of trapped and
marginally trapped surfaces in the vicinity of any dynamical
horizon. These restrictions are used to prove several uniqueness
theorems for dynamical horizons. Ramifications of some of these
results to numerical simulations of black hole spacetimes are
discussed. Finally several expectations on the interplay between
isometries and dynamical horizons are shown to be borne out.

\end{abstract}

\section{Introduction}
\label{s1}

In general relativity, the surface of a black hole is often
identified with a future event horizon $E$ \cite{swh,he}
--- the future boundary of the past of future
null infinity, $\scri^+$. This strategy is physically well
motivated because $E$ is the boundary of an interior spacetime
region from which causal signals can never be sent to the
asymptotic observers, \emph{no matter how long they are prepared
to wait.} The region is therefore `black' in an absolute sense.
Furthermore, this identification has led to a rich body of
physically interesting results such as the laws of black hole
mechanics in classical general relativity and the Hawking
evaporation of black holes in quantum field theory in curved
spacetimes. A great deal of the current intuition about black
holes comes from the spacetime geometry near event horizons of
Kerr spacetimes and linear perturbations thereof.

However, in fully dynamical situations where the non-linearity of
general relativity can not be ignored, the strategy faces serious
limitations in a number of physical contexts. These arise from two
features of event horizons. First, the notion is teleological
because one can locate an event horizon only after having access
to the spacetime geometry to the infinite future. Second, it is
`unreasonably' global: it depends on a delicate interplay between
$\scri^+$ \emph{and the entire spacetime interior.}%
\footnote{In particular, the notion is not meaningful in general
spatially compact spacetimes. Note that the gravitational
radiation theory in exact general relativity also requires
$\scri^+$. But in contrast to event horizons, in that theory the
physical notions are extracted \emph{only} from fields near
infinity.}

The first feature has direct impact, for example, on numerical
simulations of black hole mergers. In these simulations one begins
with initial data representing two well-separated black holes,
evolves them, and studies the resulting coalescence. Since the
full spacetime geometry is the \emph{end product} of such a
simulation, one  can not use event horizons to identify black
holes while setting up the initial data, nor during their
evolution. In practice, the initial black holes are identified
with disjoint marginally trapped surfaces (MTSs) and the aim is to
study how their world tubes (called marginally trapped tubes,
MTTs) evolve and merge.%
\footnote{The precise definitions of MTSs, MTTs, dynamical
horizons, etc are given in section \ref{s2.2}.}
In the analysis of dynamics, then, one is interested in how the
areas of marginally trapped surfaces grow in response to the
influx of energy and angular momentum across the MTTs, how the
`multipoles' of the intrinsic geometry of these MTTs evolve and
how they settle down to their asymptotic values. In this analysis,
the event horizon plays essentially no role. Indeed, since event
horizons can form in a \emph{flat} region of spacetime (see, e.g.,
\cite{ak3}) and grow in anticipation of energy that will
eventually fall across them, in the fully non-linear regime
changes in the geometry of the event horizon have little to do
with local physical processes. Thus, because of its teleological
nature, in numerical simulations the notion of an event horizon
has limited use in practice.

The extremely global nature of the event horizon, on the other
hand, raises issues \emph{of principle} in quantum gravity. Let us
go beyond quantum field theory in curved space time and examine
the Hawking evaporation from the perspective of full quantum
gravity. Then, it is reasonable to expect that the singularity of
the classical theory is an artefact of pushing the theory beyond
its domain of validity. Quantum gravity effects would intervene
\emph{before} the singularity is formed and the `true'  quantum
evolution may be well-defined across what was the classical
singularity. This would drastically change the global structure of
spacetime. Even if the full spacetime is asymptotically flat (with
a complete $\scri^+$), there may well be `small regions around
what was the classical singularity' in which geometry is truly
quantum mechanical and can not be
well-approximated by \emph{any} continuous metric.%
\footnote{Indeed there are concrete indications in loop quantum
gravity that this would happen \cite{ab}.}
Since there would be no light cones in this region, the notion of
the past of $\scri^+$ would become ambiguous and the concept of an
event horizon would cease to be useful. Problems continue to arise
even if we suppose that the modifications in the neighborhood of
what was a singularity are `tame'. Indeed, Hajicek \cite{ph} has
shown that one can alter the structure of the event horizon
dramatically
---and even make it disappear--- by changing the spacetime
geometry in a Planck size neighborhood of the singularity. Thus,
while the `true' spacetime given to us by quantum gravity will
have large semi-classical regions, there may not be a well-defined
event horizon. What is it, then, that evaporates in the Hawking
process? Is there an alternate way of speaking of formation and
evaporation of the black hole? One would expect that if one stayed
away from the neighborhood of what was a singularity, one can
trust classical general relativity to an excellent degree of
approximation. In these `classical' regions, there are MTSs and
MTTs. Since their area is known \cite{sh1,sh2,ak2,ak3} to grow
during collapse and decrease because of Hawking radiation, one
could use the MTTs to analyze the formation and evaporation of
black holes \cite{ab}.

Since they can be defined quasi-locally, the marginally trapped
tubes are free of the teleological and global problems of event
horizons. The above considerations suggest that, in certain
contexts, they can provide a useful quasi-local description of
black holes. Therefore, it is well worthwhile --and, in quantum
theory, may even be essential--- to understand their structure. A
decade ago, Hayward \cite{sh1,sh2} began such an analysis and over
the last four years there has been significant progress in this
direction (for a review, see \cite{ak3}). In broad terms, the
situation can be summarized as follows. If an MTT is spacelike, it
is called a \emph{dynamical horizon} (DH). Under certain
genericity conditions (spelled out in section \ref{s2.2}) it
provides a quasi-local representation of an evolving black hole.
If it is null, it provides a quasi-local description of a black
hole in equilibrium and is called an \emph{isolated horizon} (IH).
If the MTT is timelike, causal curves can transverse it in both
inward and \emph{outward} directions, whence it does not represent
the surface of a black hole in any useful sense. Therefore it is
simply called a \emph{timelike membrane} (TLM). Isolated and
dynamical horizons have provided extremely useful structures for
numerical relativity, black hole thermodynamics, quantum gravity
and mathematical physics of hairy black holes
\cite{ihprl,abl2,ak1,ak2,acs}. For instance, using Hamiltonian
methods which first provided the notions of total mass and angular
momentum at spatial infinity, it has been possible to define mass
and angular momentum of DHs \cite{bf}. There exist balance laws
relating values of these quantities on two MTSs of a DH and energy
and angular momentum fluxes across the portion of the DH bounded
by them. These notions have been useful in extracting physics from
numerical simulations (see, e.g., \cite{num1,num2,num3,ak3}). They
have also led to integral generalizations of the differential
first law of black hole mechanics, now covering processes in which
the dynamics can not be approximated by a series of infinitesimal
transitions from one equilibrium state to a nearby one. Thus, DHs
have provided a number of concrete insights in to fully non-linear
processes that govern the growth of black holes in exact general
relativity.

However, as a representation of the surface of an evolving black
hole, general DHs appear to have a serious drawback: a given
trapped region which can be intuitively thought of as containing a
single black hole may admit many dynamical horizons. While results
obtained to date would hold on all of them, it is nonetheless
important to investigate if there is a canonical choice or at
least a controllable set of canonical choices. The lack of
uniqueness could stem from two potential sources. The first is
associated with the fact that a DH $H$ comes with an assigned
foliation by MTSs. One is therefore led to ask: Can a given
spacelike 3-manifold $H$ admit two distinct foliations by MTSs? If
so, the two would provide distinct DH structures in the given
spacetime. The second question is more general: Can a connected
trapped region of the spacetime admit distinct spacelike
3-manifolds $H$, each of which is a dynamical horizon? Examples
are known \cite{bgvdb} in which the marginally trapped tubes have
more than one spacelike portions which are separated by timelike
and null portions. This type of non-uniqueness is rather tame
because the DHs lie in well separated regions of spacetime. It
would be much more serious if one could \emph{foliate} a spacetime
region by DHs. Can this happen? If there is genuine
non-uniqueness, can one introduce physically reasonable conditions
to canonically select `preferred' DHs?

The purpose of this paper is to begin a detailed investigation of
these uniqueness issues. Using a maximum principle, we will first
show that if a 3-manifold $H$ admits a DH structure, that
structure is unique, thereby providing a full resolution of the
first set of uniqueness questions. For the second set of
questions, we provide partial answers by introducing additional
mild and physically motivated conditions. In particular, we will
show that DHs are not so numerous as to provide a foliation of a
spacetime region. Some of our results have direct implications for
numerical relativity which are of interest in their own right,
independently of the uniqueness issue.

The paper is organized as follows. Section \ref{s2} is devoted to
preliminaries. We briefly recall the basics of the DH theory and
aspects of causal theory that are needed in our analysis. We also
introduce the notion of a \emph{regular} DH by imposing certain
mild and physically motivated restrictions. In section \ref{s3},
we show that the MTS-foliation of a dynamical horizon is unique.
Section \ref{s4} establishes several constraints on the location
of trapped and marginally trapped surfaces in the vicinity of a
regular DH. These are then used to obtain certain restrictions on
the occurrence of regular DHs, which in turn imply that the number
of regular DHs satisfying physically motivated conditions is much
smaller than what one might have a priori expected. In section
\ref{s5} we obtain certain implications of our main results which
are of interest to numerical relativity. Section \ref{s6}
discusses restrictions on the existence and structure of DHs in
presence of spacetime Killing vectors. We conclude in section 7
with a summary and outlook.

We will use the following conventions. By {\it spacetime} $(\calM,
g)$ we will mean a smooth, connected, time-oriented, Lorentzian
manifold equipped with a metric of  Lorentzian signature. (By
`smooth' we mean $\calM$ is $C^{k+1}$ and $g$ is $C^{k}$ with
$k\ge 2$.) The torsion-free derivative operator compatible with
$g$ is denoted by $\nabla$. The Riemann tensor is defined by
$R_{abc}{}^d W_d := 2 \nabla_{[a} \nabla_{b]}W_c$, the Ricci
tensor by $R_{ab} := R_{acb}{}^c$, and the scalar curvature by $R
:= g^{ab} R_{ab}$. We will assume Einstein's field equations
\begin{equation}
  \label{fe}
  R_{ab}- \frac{1}{2}R\, g_{ab} + \Lambda g_{ab} = 8\pi G T_{ab}.
\end{equation}
For most of our results we will assume that the null energy
condition (NEC) holds: for all null vectors $\ell$, the Ricci
tensor satisfies $R_{ab} \ell^a\ell^b \ge 0$ or, equivalently,
matter fields satisfy $T_{ab} \ell^a\ell^b \ge 0$.

We do not make an assumption on the sign or value of the
cosmological constant. For definiteness, the detailed discussion
will be often geared to the case when $\calM$ is 4-dimensional.
However, most of our results hold in arbitrary number of dimensions.
Finally, the trapped and marginally trapped surfaces are only
required to be compact; they need not be spherical.

\section{Preliminaries}
\label{s2}

\subsection{Aspects of Causal Theory}
\label{s2.1}

For the convenience of the reader, and in order to fix various
notations and conventions, we recall here some standard notions
and results from the causal theory of spacetime, tailoring the presentation to
our needs. For details, we refer the reader to the excellent
treatments of causal theory given in \cite{rp,he,on,be}. By and
large, we follow the conventions in \cite{he,on}.

Let $(\calM, g)$ be a  {\it spacetime}. We begin by recalling the
notations for futures and pasts. $I^+(p)$ (resp., $J^+(p)$), the
timelike (resp., causal) future of $p\in \calM$, is the set of
points $q\in\calM$ for which there exists a future directed
timelike (resp., causal) curve from $p$ to $q$.  Since small
deformations of timelike curves remain timelike, the sets $I^+(p)$
are always open.  However, the sets $J^+(p)$ need not in general
be closed. More generally, for $A\subset \calM$, $I^+(A)$ (resp.,
$J^+(A)$), the timelike  (resp., causal) future of $A$, is the set
of points $q\in \calM$ for which there exists a future directed
timelike (resp., causal) curve from  a point $p\in A$ to $q$.
Note, $I^+(A)= \cup_{p\in A}I^+(p)$, and hence the sets $I^+(A)$
are open.  The timelike and causal pasts $I^-(p)$, $J^-(p)$,
$I^-(A)$, $J^-(A)$ are defined in a time dual manner.

We recall some facts about global hyperbolicity. {\it Strong
causality} is said to hold at $p\in \calM$ provided there are no
closed or `almost closed' causal curves passing through $p$ (more
precisely, provided $p$ admits arbitrarily small neighborhoods $U$
such that any future directed causal curve which starts in and leaves $U$ never
returns to $U$). $(\calM, g)$ is {\it strongly causal} provided
strong causality holds at each of its points.  $(\calM, g)$ is
said to be {\it globally hyperbolic} provided (i) it is strongly
causal and (ii) the sets  $J^+(p) \cap J^-(q)$ are compact for all
$p,q \in \calM$. The latter condition rules out the occurrence of
naked singularities in $\calM$. Globally hyperbolic spacetimes are
necessarily {\it causally simple}, by which is meant that the sets
$J^{\pm}(A)$ are closed whenever $A$ is compact.   Thus, in a
globally hyperbolic spacetime,  $\d J^+(A)  = J^+(A) \setminus
I^+(A)$ for  all compact subsets $A$. This implies that, if $q \in
\d J^+(A)\setminus A$, with $A$ compact, there exists a future
directed null geodesic from a point in $A$ to $q$.  Indeed, any
causal curve  from a point $p \in A$ to $q \in  \d J^+(A)\setminus
A$ must be a null geodesic (when suitably parameterized), for
otherwise it could be deformed to a timelike curve from $p$ to
$q$, implying $q\in I^+(A)$.

Even when spacetime itself is not globally hyperbolic, one can
sometimes identify certain globally hyperbolic subsets of
spacetime, as described further below. A subset $U \subset \calM$
is said to be globally hyperbolic provided (i) strong causality
holds at each point of $U$, and (ii) the sets  $J^+(p) \cap
J^-(q)$ are compact and contained in $U$ for all $p,q \in U$.

Global hyperbolicity can be characterized in terms of the domain
of dependence. Let $S$ be a smooth spacelike hypersurface in a
spacetime $(\calM, g)$, and assume $S$ is achronal,  by which is
meant that no two points of $S$ can be joined by a timelike curve.
(A spacelike hypersurface is always {\it locally} achronal, but in
general achronality can fail in-the-large.)
$S$ will not in general be a closed subset of spacetime. (In our
applications, $S$ will correspond to a dynamical horizon, defined
in section \ref{s2.2}.) The {\it past domain of dependence} of $S$
is the set $D^-(S)$ consisting of all points $p\in \calM$ such
that every future inexendendible causal curve from $p$ meets $S$.
Physically, $D^-(S)$ is the part of spacetime to the past of $S$
that is predictable from $S$. The {\it past Cauchy horizon} of
$S$, $H^-(S)$, is the past boundary of $D^-(S)$; formally, $H^-(S)
= \{x\in \overline{D^-(S)}: I^-(x) \cap D^-(S) = \emptyset\}$.
Note, by its definition, $H^-(S)$ is a closed subset of spacetime,
and is always achronal.
We record the following elementary facts: (i) $S \subset D^-(S)$
and $D^-(S) \setminus S$ is open, (ii) $D^-(S) \cap H^-(S) =
\emptyset$, (iii) $\overline{D^-(S)} = D^-(S) \cup H^-(S)$, and
(iv) $\d D^-(S) = S \cup H^-(S)$.
The future domain of  dependence $D^+(S)$ and future Cauchy
horizon $H^+(S)$ are defined in a time-dual manner, and of course
satisfy time-dual properties.
The full domain of dependence  $D(S) = D^+(S) \cup D^-(S)$ has the
following important property:  $D(S)$ is an open globally
hyperbolic subset of spacetime.  By definition, $S$ is  a {\it
Cauchy surface} for $(\calM, g)$ provided $D(S) = \calM$.  Hence,
if $S$ is a Cauchy surface for $(\calM, g)$, $(\calM, g)$ is
globally hyperbolic.  The converse also holds, i.e., if $(\calM,
g)$ is globally hyperbolic then it admits a Cauchy surface.

\subsection{Horizons}
\label{s2.2}

In this sub-section we will summarize the definition and basic
properties of dynamical horizons that will be used in the main
body of the paper. For further details see, e.g., \cite{ak3}.

We begin by giving the precise definitions of certain notions
which were introduced in a general way in section \ref{s1}. By a
(future) \emph{trapped surface} (TS) we will mean a smooth,
closed, connected spacelike 2-dimensional submanifold of
$(\calM, g)$ both of which null normals $\ell, n$ have negative
expansions.%
\footnote{Unless otherwise stated, all null normals will be
assumed to be future pointing.}
A (future) \emph{marginally trapped surface} is a smooth, closed,
connected spacelike 2-dimensional submanifold of $(\calM, g)$ with
the property that one of its null normals, denoted $\ell$, has
zero expansion and the other, $n$, has negative expansion. While
these two notions are standard, in our discussion we will often
use a third, related notion. By a (future) \emph{weakly trapped
surface} (WTS), we will mean a  smooth, closed, connected,
spacelike 2-dimensional submanifold of $(\calM, g)$ the expansions
of both of whose null normals are nowhere positive. Finally, in
certain situations, e.g., asymptotically flat contexts, one of the
two null normals to a closed spacelike 2-manifold $S$ can be
singled out as being `outward pointing' and the other as being
`inward pointing'. For such surfaces $S$, one can introduce an
additional notion that will be useful occasionally. $S$ will be
said to be \emph{outer marginally trapped} (OMT) if the outward
null normal $\ell$ has zero expansion. Note that every MTS is WTS.
It is also OMT if the null normal with zero expansion is pointing
outward. However, there is no simple relation between weakly
trapped and outer marginally trapped surfaces. On a WTS,
expansions of both null normals are restricted but one does not
need to know which of them is outward pointing. The notion of an
OMT surface, on the other hand, restricts the expansion only of
the outward pointing normal.

A marginally trapped tube (MTT) is a smooth, 3-dimensional
submanifold of $(\calM, g)$ which admits a foliation by marginally
trapped surfaces. In numerical simulations of black holes, one
assumes that the initial data on a Cauchy surface admit an MTS,
evolves them through Einstein's equations and locates an MTS on
each subsequent constant time slice. The union of these
1-parameter family of MTSs is an important example of an MTT. If
an MTT is spacelike, it is called a \emph{dynamical horizon} (DH).
If it is timelike, it is called a \emph{timelike membrane} (TLM).
If it is null, it is called a \emph{non-expanding horizon} (NEH),
provided the spacetime Ricci tensor is such that $-R_{ab}\ell^b$
is future causal. The intrinsic `degenerate metric' $q$ on an NEH
is `time-independent' in the sense that $\mathcal{L}_\ell q =0$
for any null normal $\ell$. On physical grounds
\cite{ihprl,abl2,ak3}, one often requires that the intrinsic
connection and matter fields be also time-independent. In this
case, the NEH is called an \emph{isolated horizon} (IH). Under certain
regularity conditions, an isolated horizon provides a quasi-local
description of a black hole which has reached equilibrium, while a
dynamical horizon represents an evolving black hole. There is a
general expectation that under physically reasonable conditions,
DHs `settle down' to IHs in asymptotic future
\cite{bgvdb,num2,num3,acg}. As explained in section \ref{s1},
because causal curves can transverse a timelike hypersurface in
both `inward' \emph{and} `outward' directions, one does not
associate a TLM with the surface of a black hole even
quasi-locally.

In this paper we will deal primarily with DHs which we will denote
by $H$. A priori, it may appear that a given spacelike 3-manifold
could admit distinct foliations by MTSs. However, we will show in
section \ref{s3} that this can not happen. The MTSs of the unique
foliation of a DH will be referred to as its \emph{leaves}. More
general sections will be referred to simply as cross-sections. We
will generally work with DHs $H$ which satisfy two additional mild
conditions. A DH $H$ will be said to be \emph{regular} if i) $H$
is achronal, and ii) satisfies a `generiticity condition' that
$\sigma_{ab}\sigma^{ab} + T_{ab}\ell^a \ell^b$ never vanish on
$H$, where $\sigma_{ab}$ is the shear of the MTSs  (i.e., the
trace-free part of the projection to each MTS of $\nabla_{a}
\ell_{b}$). Since it is spacelike, every DH $H$ is automatically
locally achronal. The first regularity condition asks that it be
globally achronal. Using the spacelike character of $H$, it is
easy to show that the second (i.e., genericity) condition is
satisfied if and only if $\calL_n \tl$ never vanishes on $H$ where
$\tl$ is the expansion of $\ell$ (and $\ell$ is extended off $H$
using the geodesic equation). Furthermore, if the null energy
condition (NEC) holds, one can show that $\calL_n \tl$ is
everywhere negative. This is precisely the condition used by
Hayward \cite{sh1,sh2} in the definition of his future, outer
trapping horizon (FOTH). Thus, if the NEC holds, a DH $H$ satisfies the genericity
condition if and only if it is a FOTH. This condition serves to
single out black hole horizons from the DHs that arise in
cosmological and other contexts \cite{jmms}.

While the notions introduced so far are quasi-local and meaningful
even in spacetimes with spatially compact Cauchy surfaces, the
notion of an event horizon is well defined only in the
asymptotically flat context. Technically, it requires that the
spacetime be asymptotically flat in the sense that it has a
complete future null infinity, $\scri^+$ (for the precise
definition, see, e.g., \cite{ax}). The \emph{event horizon} $E$ is
then defined to be $E = \d I^-(\scri^+)$. Event horizons of
stationary, asymptotically flat spacetimes provide important
examples of isolated horizons. In section \ref{s4.2} and \ref{s5},
we will assume standard results \cite{swh,he} about event
horizons, such as the area theorem ($E$ has nonnegative expansion,
$\th \ge 0$), and the result that WTSs cannot meet the exterior
region $I^-(\scri^+)$.

\section{Uniqueness of the dynamical horizon structure}
\label{s3}

With the required notions at hand, we can now establish the
uniqueness of the foliation of a DH by MTSs. This uniqueness
follows easily from a geometric maximum principle closely related
to the geometric maximum principle of minimal surface theory.

Let $M$ be a spacelike hypersurface in a spacetime $(\calM, g)$,
with induced metric $q$. Let $\hat\tau$ denote the future pointing
unit normal of $M$.  Given two surfaces $S_i$, $i = 1,2$, in $M$,
let ${\hat r}_i$ denote the `outward pointing' unit normal of
$S_i$ within $M$, and  let $\theta_i$ denote the null expansion
scalar of $S_i$ with respect to the outward null normal vector
${\ell_i} = {\hat r}_i + {\hat \tau}$.

\begin{prop}\label{dhmaxprin} Suppose $S_1$ and $S_2$ meet
tangentially at a point $p$ (with  `outsides' compatibly oriented,
i.e., $\ell_1 = \ell_2$ at $p$), such that near $p$, $S_2$ is to
the outside of $S_1$. Suppose further that the null expansion
scalars satisfy, $\theta_1 \le 0 \le \theta_2$.  Then $S_1$ and
$S_2$ coincide near $p$ (and this common surface has zero null
expansion, $\theta = 0$).
\end{prop}

\noindent{\it Remarks:}  Since the sign of the null expansion
scalar is invariant under the scaling of the associated null
field, Proposition \ref{dhmaxprin} does not depend on the
particular normalizations chosen for the null fields $\ell_1$ and
$\ell_2$.  The proof of Proposition~\ref{dhmaxprin} is similar to
the proof of the maximum principle for smooth null hypersurfaces
obtained in \cite{gg1}, which  may be consulted for additional
details (see also \cite{agh,gg2} for related results and further background).

\medskip
\proof   The proof is an application  of the strong maximum
principle for second order quasi-linear elliptic PDEs~\cite{gt,agh}.
We begin by setting up the appropriate analytic framework. Let $S$
be a surface in $(M,q)$,  with induced metric $\tilde q$, and with
outward null normal field $\ell= \hat r + \hat\tau$. Let
$\bar\theta$ denote {\it minus} the  expansion scalar of $S$ with
respect to $\ell$, $\bar \theta = -\theta$. From the definition of
the null expansion scalar we have,
\beq\label{theta} \bar \theta = - {\tilde K}
- {\tilde q}^{ab}\D_a\hat\tau_b \, ,
\eeq
where $\tilde K := {\tilde q}^{ab}D_a\hat r_b$ is the mean
curvature of $S$ in $M$.
We need to express the above equation  in terms of relevant
differential operators.

Fix a point $p$ in $S$, and a surface $\S$ through $p$ tangent to
$S$ at $p$.  Introduce Gaussian normal coordinates along $\S$ so
that in a neighborhood $U\subset M$ of $p$ we have
\beq\label{gauss} U = (-\e, \e) \times \S\,,\quad ds^2 = d\xi^2 +
q_{ij}(t,x)dx^idx^j   \, . \eeq
where $x =(x^1,x^2)$ are coordinates in $\S$ (and $\d/\d \xi$
points to the outside of $S$ at $p$).  By adjusting the size of
$\S$ and $U$ if necessary, $S$ can be expressed in $U$ as a graph
over $\S$, i.e. there is a smooth function $u(x)$ on $\S$, such
that $S =$ graph\,$u = \{(u(x),x):x\in \S\}$.  Let $\bar\Theta(u)$
denote minus the expansion scalar of $S = {\rm graph}\, u$.  Then
from (\ref{theta}) one obtains,
\beq\label{Theta} \bar\Theta(u) = - \mathcal{H}(u) +  b(x,u,\d
u)\,, \eeq
where $\mathcal{H}$ is the mean curvature operator (for graphs
over $\S$), and $b(x,u,\d u)$ is a term involving $u$ up to first
order which comes  from the term involving the extrinsic curvature
of $M$  in (\ref{theta})  (cf., \cite{gg1} for this computation in
the closely related situation in which $M$ is timelike).

$\mathcal{H}$ is a second order quasi-linear operator, which, with
respect to the Gaussian coordinates (\ref{gauss}),  can be written
as
\beq\label{mean} \mathcal{H}(u)  = -  a^{ij}\frac{\partial^2
u}{\partial x^i\partial x^j} + \mbox{ lower order terms in } u\,,
\eeq
where $a^{ij} = a^{ij}(x,u,\partial u)$ is such that the matrix
$[a^{ij}]$ is symmetric and positive definite.  In fact, a
computation shows that $a^{ij} = w\, \gamma^{ij}$, where
$\g_{ij}$, $1\le i,j \le 2$, are the components of the induced
metric $\tilde q$ on $S$ with respect to the coordinates
$x_1,x_2$, and   $w = \left(1+ q^{ij}(u(x),x)\, \frac{\d u}{\d
x^i}\frac{\d u}{\d x^j}\right)^{-\frac12}$.
Combining (\ref{Theta}) and (\ref{mean}), we obtain,
\beq \bar\Theta (u) =  a^{ij} \frac{\partial^2 u}{\partial
x^i\partial x^j} + \mbox{ lower order terms in } u \,, \eeq
where $a^{ij}$ is as described above.  We conclude that $\bar
\Theta$ is a second order quasi-linear elliptic operator.

Now focus on the surfaces $S_1$ and $S_2$ of the proposition.
Introduce Gaussian coordinates as in (\ref{gauss}), where $\S$ is
a surface in $M$ tangent to $S_1$ and $S_2$ at $p$.  Again,
adjusting the size of $\S$ and $U$ if necessary, $S_1$ and $S_2$
can be expressed in $U$ as graphs over $\S$, $S_i = {\rm
graph}\,u_i$, $u_i \in C^{\infty}(\S)$, $i =1,2$, with the graph
of $S_1$ lying below the graph of $S_2$. The assumptions of
Proposition \ref{dhmaxprin} become,
\ben \item[(1)] $u_1 \le u_2$ on $\S$ and  $u_1(p) = u_2(p)$, and
\item[(2)] $\bar\Theta(u_2) \le  0 \le \bar\Theta(u_1)$\, on $\S$.
\een
By the strong maximum principle for second order quasi-linear
elliptic operators~\cite{gt,agh}, $u_1 = u_2$, and hence $S_1 = S_2$
near $p$. \qed

The uniqueness of the foliation of a DH by MTSs is an immediate
consequence of the following corollary.  To fix notation,
let  $H = \bigcup_{R \in (R_1,R_2)} S_R$ denote a dynamical horizon
foliated by MTSs $S_R$, $R_1 < R < R_2$, where
$R$ is the area radius.

\begin{cor}\label{cor1}
Let $S$ be a  WTS in a DH $H = \bigcup_{R \in (R_1,R_2)} S_R$.
Then $S$ coincides with one of the $S_R$'s.
\end{cor}

\proof The area radius $R$ achieves a maximum, $R_0$, say, on $S$
at some point $p\in S$. With respect to the outward direction
(direction of increasing $R$) near $p$, $S$ satisfies $\theta \le
0$, and $S_{R_0}$ satisfies $\theta =0$.   Moreover, by the choice
of $R_0$, $S$ is contained in the region $R \le R_0$. It follows
that $S$ and $S_{R_0}$ satisfy the hypotheses of Proposition
\ref{dhmaxprin}, and thus, must agree near $p$.  By similar
reasoning, it follows that the set of points in $S$ (or $S_{R_0}$)
where $S$ and $S_{R_0}$ agree is open, as well as closed, and
hence they must agree globally, $S = S_{R_0}$. \qed

\section{Constraints on the occurrence of MTSs and DHs}
\label{s4}

In this section we obtain some constraints on the occurrence of
MTSs and DHs in the {\it vicinity} of a given regular DH. We begin
by considering results about MTSs, from which results about DHs
will follow straightforwardly.

\subsection{Constraints on the occurrence of MTSs}
\label{s4.1}

Let $H = \bigcup_{R \in (R_1,R_2)} S_R$ denote a dynamical horizon
foliated by MTSs $S_R$, $R_1 < R < R_2$, where
$R$ is the area radius. Let $H_{R > R_0}$ denote
the part of $H$ for which $R>R_0$, i.e., $H_{R > R_0} = \bigcup_{R
\in (R_0,R_2)} S_R$.

Our first and most basic result rules out the existence of weakly
trapped surfaces (WTSs) in the past domain of dependence $D^-(H)$
of a DH $H$ in a spacetime satisfying the null energy condition.

\begin{thm}\label{basic1}  Let $H$ be a regular  DH in a
spacetime $(\calM, g)$ satisfying the NEC. Then  there are no
weakly trapped surfaces (WTSs) contained in $D^-(H) \setminus H$.
\end{thm}

Theorem \ref{basic1} implies that one can not foliate a region of
space-time with regular DHs even locally.%
\footnote{Even without the genericity condition, such a foliation
could exist only under very special circumstances, e.g., in
space-times in which all curvature scalars vanish \cite{jmms}. In
the absence of the genericity condition, it can be shown that if a
WTS is contained in $D^-(H) \setminus H$, it must be joined to a
MTS of $H$ by a null hypersurface of vanishing expansion and
shear.}
However, it does not rule out the existence of a WTS $\S$ not
entirely contained in $D^-(H) \setminus H$. The next two results
allow for this possibility and then restrict the {\it location} of
such a WTS. In these results, it is assumed that $\S$ does not
meet $H^-(H)$, the past Cauchy horizon of $H$. Even when $\S$ does
not lie entirely in $D^-(H)$, this assumption can be met (for
example, if $\S$ is contained in $D^-(H) \cup J^+(H)$).

\begin{thm}\label{basic2}
Let $H$ be a regular  DH in a spacetime $(\calM, g)$ satisfying
the NEC. Suppose $\S \subset \calM \setminus H^-(H)$ is a WTS that
meets $H$ in a fixed leaf $S_{R_0}$.   Then $\S \cap D^-(H)$
cannot meet the causal past of  $H_{R > R_0}$.
\end{thm}

In fact, Theorems \ref{basic1} and \ref{basic2} are immediate
consequences of the following more general result, which allows
for a more general intersection of  the WTS with $H$.

\begin{thm}\label{basic3}
Let $H$ be a regular  DH in a spacetime $(\calM, g)$ satisfying
the NEC. Let $\S$  be a WTS that does not meet $H^-(H)$.  If $\S
\cap H \ne \emptyset$, set
\beq\label{sup} R^* = \sup_{x\in \S \cap H} R(x)\, ; \eeq
otherwise set $R^* = R_1$.  Then $\S \cap D^-(H)$ cannot meet the
causal past of  $H_{R > R^*}$.
\end{thm}

We remark that, since $H$ is closed in $\calM \setminus H^-(H)$
and $\S$ doesn't meet $H^-(H)$, $\S \cap H$ is compact, so that
the sup in (\ref{sup}) is achieved, and hence, $R^* < R_2$ (see figure~1).

\medskip
\noindent

For the reader's convenience, we will first present a rough, heuristic
argument that captures the main idea of the proof. Let $\S_0$ be
the portion of $\S$ in $D^-(H)$. Consider the intersection with
$H$ of the two null hypersurfaces generated by the future directed
null geodesics issuing orthogonally from $\S_0$. One of the two,
call it $\calN$, will meet $H$ at a point $q$ having a largest
$R$-value in this intersection. If the conclusion of the theorem
does not hold then this largest $R$-value, call it $R_0$, will be
strictly greater than $R^*$, $R_0 >R^*$.  This implies that $q$ is
away from  $\S_0 \cap H$, and hence is in the interior of $\calN$.
The NEC and the fact that $\S$ is (weakly) trapped imply that
$\calN$ has negative expansion at $q$.  By a `barrier argument',
this forces the expansion of the null hypersurface generated by
$\ell$ along the MTS $S_{R_0}$ to be negative at $q$, $\tl(q) <
0$.  But this contradicts the fact that $\tl$ vanishes along
$S_{R_0}$. While containing a number of inaccuracies, this
description captures the spirit of the proof.

We now proceed to a rigorous proof.  For technical reasons the
barrier argument alluded to above will take place at a point
slightly to the past of $H$.

\begin{figure}
\begin{center}
\includegraphics[height=2in]{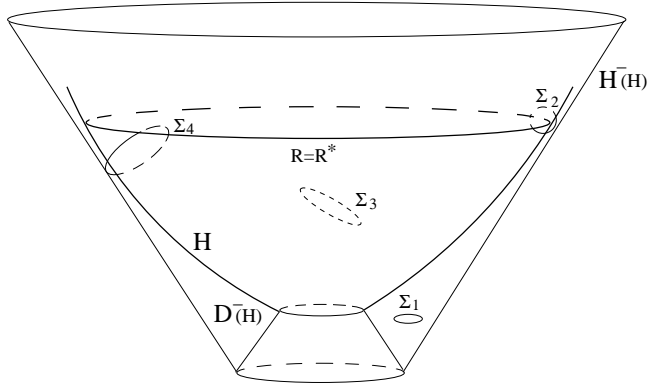}
\caption{\small Examples of situations ruled out by Theorem
\ref{basic3} as well as of those that are not. Here $H$ is the
dynamical horizon, $H^-(H)$, its past Cauchy horizon, and
$D^-(H)$, its past domain of dependence. The compact surface
$\S_1$ lies in $D^-(H)\setminus H$ while $\S_3$ lies entirely to
the future of $H$. Compact surfaces $\S_2$ and $\S_4$ intersect
$H$ and $R^*$ is the largest value of $R$ on their intersections
with $H$. $\S_2$ meets the causal past of $H_{R > R^*}$. $\S_4$
meets $H^-(H)$. The theorem implies that $\S_1$ and $\S_2$ can not
be WTSs but does not rule out the possibility that $\S_3$ and
$\S_4$ are WTSs.} \label{fig1}
\end{center}
\end{figure}

\medskip
\noindent
{\it Proof of Theorem \ref{basic3}:} Set $\S_0 := \S \cap D^-(H)$;
since $\S$ does not meet $H^-(H)$ we have, $\S_0 = \S \cap (D^-(H)
\cup H^-(H)) = \S \cap \overline{D^-(H)}$, and hence $\S_0$ is
compact.

The full domain of dependence of $H$, $D(H) = D^+(H) \cup D^-(H)$,
is an open globally hyperbolic subset of spacetime, and $H$ is a
Cauchy surface for $D(H)$, viewed as a spacetime in its own right.
Thus, by restricting to $D(H)$, we may assume without loss of
generality that spacetime $(\calM, g)$ is globally hyperbolic, and
$H$ is a Cauchy surface for it.

Assume, contrary to the stated conclusion, that $\S_0$ meets the
causal past of $H_{R > R^*}$. Then, $J^+(\S_0)$ meets $H_{R >
R^*}$. Since $H$ is Cauchy and $\S_0$ is compact, $A := J^+(\S_0)
\cap H$ is a compact subset of $H$.  Let $R_0 = \sup_{x\in A}
R(x)$. Since $A$ is compact this sup is achieved for some point
$q\in A$ and hence $R(q) = R_0$. Clearly, $q$ must lie on the
boundary of $A$ in $H$, which implies that $q\in \d J^+(\S_0) = J^+(\S_0)
\setminus I^+(\S_0)$. Since $J^+(\S_0)$ meets $H_{R > R^*}$,  it
follows that $R_0 > R^*$, and hence $q$ is not on $\S_0$.  Thus,
there exists a future directed null geodesic segment $\eta$
extending from a point $p \in \S_0 \setminus H$ to $q$. Note that
$\eta$ must meet $\S_0$ orthogonally at $p$, for otherwise, by a
small deformation of $\eta$, there would exist a timelike segment
from a point on $\S_0$ to a point on $\eta$, which would imply
that $q$ is in $ I^+(\S_0)$. For similar reasons, there can be no
null geodesic, distinct from $\eta$, joining a point of $\S_0$ to
a point of $\eta$ prior to $q$, and there can be no null focal
points to $\S_0$ along $\eta$ prior to $q$ \cite{on}.   Then, by
standard \cite{on,be} results, we know that the congruence of
future directed null geodesics near $\eta$, issuing orthogonally
to $\S_0$ from points on $\S_0$ near $p$, forms a smooth null
hypersurface in a neighborhood of $\eta \setminus \{q\}$. (We need
to avoid $q$, due to the possibility that it may be a focal point,
and hence a point where smoothness may break down.) Call this null
hypersurface ${\bf N_1}$ (see figure~2).

Now consider the smooth null hypersuface ${\bf N_2}$, defined in a
neighborhood of $S_{R_0}$,  generated by the null geodesics
passing through the points of $S_{R_0}$ in the direction of the
outward null normal $\ell$ along $S_{R_0}$. We wish to compare
${\bf N_1}$  and ${\bf N_2}$ near $q$.
To make this comparison, we first observe that $\eta$ meets
$S_{R_0}$ orthogonally at $q$.  For if  it didn't, there would
exist a timelike curve from a point of $\eta$ to a point  $y\in
S_{R_0}$. Hence, $y\in I^+(\S_0)$ would be an interior point of
$A$ satisfying $R(y) = R_0$.  But this would mean that there are
points $x$ in $A$ with $R(x) > R_0$, contradicting the choice of
$q$. Thus, $\eta$ meets $S_{R_0}$ orthogonally, and hence,
$\eta'|_q$, the tangent vector to $\eta$ at $q$,  points  in the
direction either of the outward null normal $\ell|_q$ or the
inward null normal $n|_q$.  But note, in the latter case, there
would exist a timelike curve from a point on $\eta$ slightly to
the past of  $q$ to a point $z\in H$ satisfying, $R(z) > R_0$,
again contradicting the choice of $q$.

\begin{figure}
\begin{center}
\includegraphics[height=2in]{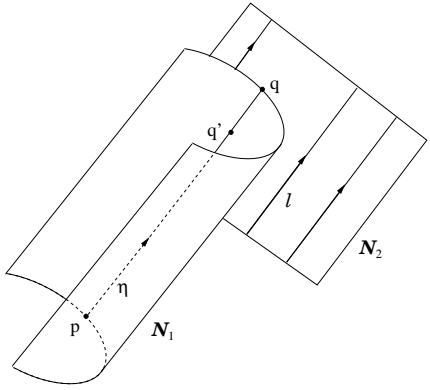}
\caption{\small Null hypersurfaces ${\bf N_1}$ and ${\bf N_2}$.
${\bf N_1}$ is generated by future directed null geodesics issuing
orthogonally to $\S_0$ from points near $p$ in $\S_0$. ${\bf N_2}$
is generated by null geodesics along $\ell$, passing through
points of $S_{R_0}$ near $q$. The segment near $q'$ of the null
geodesic $\eta$ lies on both null surfaces.} \label{fig2}
\end{center}
\end{figure}

We conclude from the preceding paragraph that,  at $q$, the
tangent vector $\eta'$   points in the same direction as
$\ell|_q$; by choosing the affine parameter of $\eta$ suitably, we
may assume $\eta'|_q = \ell|_q$.  It follows that, near,  but
prior to $q$, $\eta$ is a common null generator of ${\bf N_1}$ and
${\bf N_2}$, and ${\bf N_1}$ and ${\bf N_2}$ are tangent along
$\eta$. Choose a point $q'$ on $\eta$ slightly to the past of $q$;
we will be a bit more specific about the choice of $q'$ in a
moment. For $i=1,2$, let $\theta_i$ denote the expansion of the
null generators of ${\bf N_i}$, as determined by the tangent null
vector field $\ell_i$; $\ell_1$ and $\ell_2$ may be scaled so as
to agree along $\eta$.  We want to compare $\th_1$ and $\th_2$ at
$q'$.  We know that ${\bf N_1}$ and ${\bf N_2}$ are tangent at
$q'$. Furthermore, in a suitably chosen  small neighborhood of
$q'$, ${\bf N_1}$ must lie to the future side of ${\bf N_2}$.
Indeed, if this were not the case, there would exist a point $w'$
on ${\bf N_2}$ which is in the timelike future of a point on ${\bf
N_1}$. Following the generator of ${\bf N_2}$  through $w'$ to the
future, we obtain a point $w \in S_{R_0}$ which is in the interior
of $A$, again contradicting the choice of $q$.

From the fact that ${\bf N_1}$ and ${\bf N_2}$ are tangent at
$q'$, and that near $q'$, ${\bf N_1}$ lies to the future side of
${\bf N_2}$, it follows that the generators of ${\bf N_1}$ diverge
at least as much as the generators of ${\bf N_2}$ at $q'$, i.e.,
one must have the inequality, \beq\label{ineq} \theta_1(q') \ge
\theta_2(q') \,. \eeq Since $\S$ is a WTS, we know that
$\theta_1(p) \le 0$. Moreover, Raychaudhuri's equation for a null
geodesic congruence, in conjunction with the null energy
condition, implies that $\calL_{\ell_1}(\theta_1) \le 0$ along
$\eta$, i.e. $\theta_1$ is nonincreasing to the future along
$\eta$.  This implies that $\theta_1(q') \le 0$, and hence by
(\ref{ineq}) that $\theta_2(q') \le 0$.   On the other hand, since
$S_{R_0}$ is an MTS, $\theta_2(q) = \tl |_{q} = 0$.  Further,
since $n-\ell$ is tangent to $H$, we have $ \calL_{n-\ell}(\tl)
\equiv 0$.  Hence, $\calL_{-\ell_2}(\th_2)|_q =
\calL_{-\ell}(\tl)|_q = -\calL_{n}(\tl)|_q +
\calL_{n-\ell}(\tl)|_q =  -\calL_{n} (\tl)|_q$.  Thus, the
genericity condition implies that $\calL_{-\ell_2}(\th_2)|_q > 0$.
It follows that if  $q'$ is taken sufficiently close  to $q$,
$\th_2(q')$ will be strictly positive, and we arrive at a
contradiction.~\qed

By arguments similar to those used to prove Theorem \ref{basic1},
one may obtain the time-dual conclusion of that theorem that there
are no {\it past} weakly trapped surfaces contained in $D^+(H)
\setminus H$. This result is of physical interest in spherically
symmetric space-times because it implies that every round sphere
in $D^+(H)\setminus H$ is (future) trapped. However, we are now
considering general space-times and our present task is to
restrict the \emph{location} of (future) MTSs in the vicinity of
$H$, for which the exact time-dual can not be directly used.

\begin{figure}
\begin{center}
\includegraphics[height=1.5in]{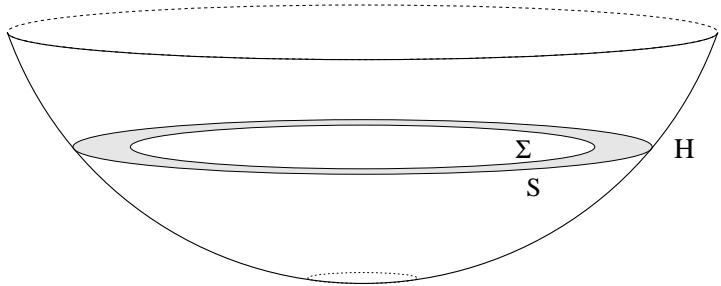}
\caption{\small $H$ is a dynamical horizon. Theorem \ref{dual}
implies that a 2-surface like $\S$ can not be marginally trapped.}
\label{fig3}
\end{center}
\end{figure}

Nevertheless, there is a partial time dual that does apply. To
formulate it, we need first to introduce a couple of definitions.
By a {\it cross-section} of a DH $H$ we mean a compact surface $S
\subset H$ that is a graph, $R = R(x)$, over some  fixed MTS
$S_{R_0}$. Then it makes sense to talk about points or sets in $H$
that lie {\it above} or {\it below} $S$; e.g., $A \subset H$ lies
below $S$ provided $A$ is contained in the region $\{ R \le
R(x)\}$, etc.. We will say that a MTS $\S$ in $D^+(H)\setminus H$
\emph{can be properly joined to} $H$ if there exist a
cross-section $S$ of $H$ and a spacelike, compact hypersurface $M
\subset D^+(H)$ with boundary $\S \cup S$ such that: i) $M \cap H
= S$, ii) $J^-(M) \cap H$ lies below $S$, and iii)  the projection
onto $M$ of the null normal $\ell$ to $\S$ with zero expansion
points towards $S$ everywhere on $\S$ (see figure \ref{fig3}).

The following result is then the desired  partial time dual of
Theorem \ref{basic1}.

\begin{thm}\label{dual}  Let $H$ be a regular  DH in a spacetime
$(\calM, g)$ satisfying the NEC. Then there does not exist an MTS
$\S$ in $D^+(H) \setminus H$ that can be properly joined to $H$.
\end{thm}

\medskip
\noindent {\it Comments on the proof:} The proof is {\it
essentially} the time dual of the proof of Theorem~\ref{basic3}.
Assuming the existence of such an MTS $\S$, we set $A = J^-(M)
\cap H$, and let $R_0 = \inf_{x \in A} R(x)$.  We choose $q \in H$
to be a point that realizes this infimum.  It follows that $q
\in \d J^-(M) \setminus M$, and since $\d J^-(M) = J^-(M)
\setminus I^-(M)$, we obtain  a past directed null geodesic $\eta$
from a point $p \in M$ to  $q$.  By arguments similar to those
used before, in order to avoid $q \in I^-(M)$ it must be that (1)
$p \in \S$, (2) $\eta$ meets $\S$ orthogonally at $p$, and (3)
$\eta'$ points in the direction of $-\ell$ (rather than $-n$) at
$p$.   The main role of the `properly joined to $H$' condition is
to insure that this third condition holds. Since $-\ell$ has
nonpositive (in fact zero)
expansion along $\S$, the proof may now proceed in precisely the
same (but time-dual) manner as the proof of Theorem \ref{basic3}:
We introduce the null hypersurfaces ${\bf N_1}$ and ${\bf N_2}$,
the former generated by the congruence of past directed orthogonal
null geodesics near $\eta$, and the latter generated by $\ell$
along $S_{R_0}$, and obtain a contradiction by examining the
expansions of these null hypersurfaces at a point $q'$ on $\eta$
slightly  to the future of~$q$.  In the present situation, ${\bf
N_1}$ lies to the {\it past} of ${\bf N_2}$ near $q'$ (at which
they are tangent).\qed

\noindent {\it Remark:} We note, as follows from the proof, that
Theorem \ref{dual} remains valid without any condition on the
expansion of the other null normal $n$ of the MTS $\S$.

\subsection{Constraints on the occurrence of DHs}
\label{s4.2}

Theorem \ref{basic1} leads immediately to statements about the
existence and location of dynamical horizons.

\begin{cor}\label{weave}
Let $H$ be a regular  DH in a spacetime $(\calM, g)$ satisfying
the NEC. If $H'$ is another  DH contained in $D(H)$, $H'$ cannot
lie strictly to the past of $H$. If $H'$ `weaves' $H$ (so that
part of it lies to the future and part of it lies to the past of
$H$), $H'$ cannot contain an MTS which lies strictly to the past
of $H$.
\end{cor}

\proof If either conclusion failed there would exist a WTS in
$D^-(H) \setminus H$, contradicting Theorem \ref{basic1}. \qed

\noindent We note that this corollary remains valid if $H'$ is
merely a MTT.

There is a general expectation that under `favorable'
circumstances a marginally trapped tube that forms during a
gravitational collapse becomes spacelike ---i.e., a DH--- soon
after collapse, and asymptotically approaches the event horizon.
This is based on some explicit examples such as the Vaidya
spacetimes \cite{ak2,ak3}, general analytical arguments
\cite{bgvdb}, numerical simulations of the collapse of a spherical
scalar field \cite{acg,bgvdb}, and of axi-symmetric collapse of
rotating neutron stars \cite{num2,num3}. From a physical and
astrophysical viewpoint, it is these dynamical horizons which
asymptote to the event (or, more generally, isolated) horizons in
the distant future that are most interesting. We now consider this
situation.

Thus, for the following, we assume spacetime $(\calM, g)$ is
asymptotically flat, i.e. admits a complete $\scri^+$ \cite{ax},
and has an event horizon $E$, for which standard
results \cite{swh,he} are assumed to hold, cf., section
\ref{s2.2}. We have in mind here, and the next section, {\it
dynamical} black hole spacetimes. (Results concerning stationary
black holes will be considered in Section 6.)  In view of this,
and to avoid certain degenerate situations, we will assume that
\emph{each cross section of the event horizon has at least one
point at which $\th > 0$}. We say that a DH $H$ {\it  is
asymptotic to the event horizon $E$} if there exists an achronal
spacelike hypersurface $V$ meeting $H$ in some MTS $S$ such that,
in the portion of spacetime to the future of $V$, the Cauchy
horizon of $H$ coincides with $E$, $H^-(H) \cap I^+(V) = E \cap
I^+(V)$.  We refer to $I^+(V)$ as the \emph{asymptotic region}.

In this setting, Corollary \ref{weave} may be formulated as follows.

\begin{cor}\label{weave2}
Let $(\calM, g)$ be a spacetime obeying the NEC and admitting an
event horizon $E$. Let $H$ be a regular DH  asymptotic to $E$.  If
$H'$ is another  DH contained in the asymptotic region, $H'$
cannot lie strictly to the past of $H$. If $H'$ `weaves' $H$ (so
that part of it lies to the future and part of it lies to the past
of $H$), $H'$ cannot contain an MTS which lies strictly to the
past of $H$.
\end{cor}

\proof Let $\S$  be an MTS of $H'$ contained in $I^-(H) \cap
I^+(V)$.  In view of Corollary~\ref{weave}, it is sufficient to
show that $\S \subset D^-(H)$.  We show first that $\S \subset
D^-(H) \cup H^-(H)$. Let $p$ be a point on $\S$, and suppose $p
\notin D^-(H) \cup H^-(H)$. Since $p \in I^-(H)$, there exists a
future directed timelike curve $\g$ from $p$ to $H$.  In order to
reach $H$, $\g$ must cross $H^-(H)$ at a point $q$, say.  Since
$\S$ is in the asymptotic region, we must have $q \in E = \d
I^-(\scri^+)$.  Hence $p \in I^-(E) \subset I^-(\scri^+)$.   But
as a WTS, $\S$ cannot meet the exterior region $I^-(\scri^+)$.
Thus, $\S \subset D^-(H) \cup H^-(H)$.

Suppose $\S \subset I^+(V)$ touches $H^-(H) \cap I^+(V)= E \cap
I^+(V)$. Then,  $\S$ lies inside the event horizon $E$, and
touches it at some point. $\S$ has expansion $\th \le 0$ with
respect to its future null normals, and $\S_E$, the cross section
of the event horizon obtained as the intersection of $E$ with
$H'$, has expansion $\th \ge 0$ with respect to its future,
outward null normal.   The maximum principle Proposition \ref{dhmaxprin}
then implies that $\S$ and $\S_E$  coincide.   This forces the
cross section $\S_E$ to have vanishing expansion, contrary to
assumption.   It follows that $\S \subset D^-(H)$, as was to be
shown.  \qed

\section{DHs and Numerical Relativity}
\label{s5}

This section is intended to be of more direct relevance to
numerical relativity.  We will begin by formalizing the strategy
used in black hole simulations to locate MTTs at each time step of
the evolution.

Given a DH $H$, let $\{M_t: t_1 < t < t_2\}$ be a foliation by
(connected) spacelike hypersurfaces of an open set $U\subset\calM$
containing $H$, given by a time function $\calT$ on $U$ (i.e. $M_t
= \{\calT\! = t\}$).  Suppose that each $M_t$ meets $H$
transversely in a unique MTS $S_R$ such that the area radius $R =
R(t)$ is an increasing (rather than decreasing) function of $t$.
Suppose further that $S_R$ separates $M_t$ into an  `inside' (the
side into which the projection of $n$ points) and an `outside'
(the side into which the projection of $\ell$ points).  By these
conditions, the outside  of $S_R$ in $M_t$ lies locally to the
past of $H$,  and the inside lies locally to the future of $H$.
Under this set of circumstances we say that $H$ is \emph{generated
by the spacelike foliation} $\{M_t\}$.

\begin{figure}
\begin{center}
\includegraphics[height=1.5in]{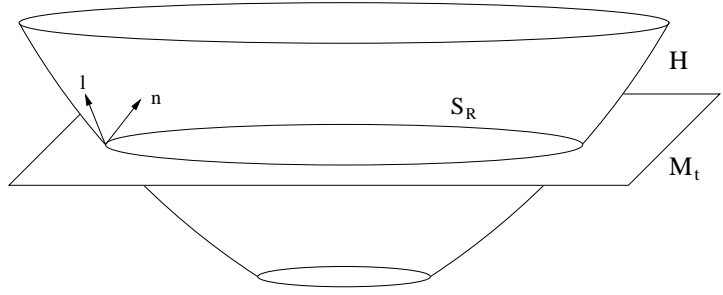}
\caption{\small Illustration of a dynamical horizon $H$ which is
generated by a foliation $M_t$. The figure shows one leaf of the
foliation which intersects $H$ in a MTS $S_R$ of area radius $R$.
$\ell$ is the outward pointing null normal to the MTSs and $n$ the
inward pointing one.} \label{fig4}
\end{center}
\end{figure}

The following result establishes conditions under which the MTSs
of a DH are `outermost'.

\begin{thm}\label{outermost} Let $H$ be a regular  DH in a spacetime
$(\calM, g)$ satisfying the NEC. Suppose $H$ is generated by a
spacelike foliation $\{M_t\}$ in $\calM \setminus H^-(H)$. Let $S=
S_{R_0}$ be the MTS of $H$ determined by the slice $M =  M_{t_0}$.
Then $S$ is the outermost WTS  in $M$.  That is, if $\S$ is any
other WTS in $M$, $\S$ does not meet the region of $M$ outside of
$S$.
\end{thm}

An example in \cite{bgvdb} shows that
the MTSs of a DH $H$  can fail to be
outermost in the sense of
Theorem \ref{outermost}, if the spacelike foliation $\{M_t\}$ is
allowed to meet $H^-(H)$.

\medskip
\proof  $M$ decomposes into the disjoint union, $M = \calI \cup S
\cup \calO$, where $\calI$ is the inside of $S$ and $\calO$ is the
outside. We want to show $\S \subset \calI \cup S$.  Suppose, to
the contrary $\S$ meets $\calO$.   Near $S$, $\calO$ is to the
past of $H$, and hence meets $D^-(H)$.  It follows that $\calO
\subset D^-(H)$. Indeed,  if  $\calO$ were to leave $D^-(H)$ it
would have to intersect $\d D^-(H) = H \cup H^-(H)$. Since $\calO$
does not meet $H$, it must intersect $H^-(H)$. But we have assumed
the entire foliation misses $H^-(H)$.

Thus, $\calO \subset D^-(H)$.  By Theorem \ref{basic1}, $\S$
cannot be entirely contained in $\calO$.  It follows that $\S$
meets $H$ in the fixed leaf $S= S_{R_0}$. By Theorem \ref{basic2},
$\S \cap D^-(H)$ cannot  meet the causal past of $H_{R > R_0}$.
Choose a point $p \in \S \cap \calO$. Since $p \in D^-(H)
\setminus H$, there exists a future directed timelike curve $\g$
from $p$ to a point $q\in H$.  By choosing $p$ sufficiently close
to $S$, we can ensure that $\g$ is entirely contained in the
foliated region. Then, since $t$, the label for the foliation
$\{M_t\}$, increases along $\g$, and  $R$ is an increasing
function of $t$,  it follows that $R > R_0$ at $q$, a
contradiction.   \qed

Theorem \ref{outermost}  implies, in particular, that DHs determined by a
spacelike foliation $\{M_t\}$ can not `bifurcate', i.e., there can
not be two distinct DHs determined by the same spacelike foliation
coming out of a common MTS.  More precisely, we have the following.

\begin{cor}\label{equal0}  Let $H_1$ and $H_2$ be DHs generated by a spacelike
foliation $\{M_t : t_1 < t <t_2\}$ in a spacetime $(\calM,g)$ satisfying the
NEC.  If, for some $t_0 \in (t_1,t_2)$, the MTS of $H_1$ and of $H_2$, respectively,
determined by  the time slice $M_{t_0}$ agree, then $H_1 = H_2$.
\end{cor}

\proof   For the purposes of this proof, let $S_1(t)$ and $S_2(t)$
denote the MTSs of $H_1$ and $H_2$, respectively, in the time
slice $M_t$.  We are assuming that $S_1(t_0) = S_2(t_0)$, and we want to
show that $S_1(t) = S_2(t)$ for all $t \in (t_1,t_2)$.  Shrink
the foliation $\{M_t\}$ to a neighbohood $U\subset D(H_1)$ of $H_1$.
Since $S_2(t_0) = S_1(t_0)$,
there exists $\e > 0$ such that $S_2(t) \subset U$
for all  $t \in (t_0 - \e,t_0 + \e)$.  Then,
Theorem \ref{outermost} applied to $H_1$ implies that
$S_2(t) \subset \calI_1(t) \cup S_1(t)$ where $\calI_1(t)$ is the
inside of $S_1(t)$ in $M_t$, for all $t \in (t_0 - \e,t_0 + \e)$.
It follows that  $H_2$ lies to the future side of $H_1$ near $S_1(t_0)$.
This in turn implies that $H_1$ lies to the past side of $H_2$
near  $S_2(t_0)$.  As a consequence we have that $S_1(t)
\subset \calO_2(t) \cup S_2(t)$, where  $\calO_2(t)$ is the
outside of $S_2(t)$ in $M_t$, for all $t \in (t_0 - \e,t_0 + \e)$.
But, by Theorem~\ref{outermost}, now applied to $H_2$,
$S_2(t)$ is outermost in $M_t$. We conclude that $S_1(t) = S_2(t)$
for all $t \in (t_0 - \e,t_0 + \e)$.

Let $A= \{t \in (t_1,t_2): S_1(t) = S_2(t)\}$.  The argument above shows
that $A$ is open in $(t_1,t_2)$.  Since it is also clearly closed and nonempty,
we have $A = (t_1,t_2)$.  Hence $H_1 = H_2$. \qed

\medskip
\noindent {\it Remark:}  In the setting and notation of Theorem
\ref{outermost}, it is possible to show that the MTS $S$ is also
outermost among outer marginally  trapped (OMT) surfaces in $M$
{\it homologous} to $S$.  For example, by arguments similar, but
time-dual, to those used to prove Theorem \ref{dual} (cf., also
the remark after the proof), one may show that there can be no OMT
surface  in $M$ lying outside of $S$ that, together with $S$,
bounds a region in $M$.  (Such a surface would be {\it properly
joined} to $S$ in a sense similar to that used in Theorem
\ref{dual}.)

\medskip

In numerical simulations one is interested in foliations $M_t$
which start out at spatial infinity, pass through the Cauchy
horizon $H^-(H)$ and meet $H$ in MTSs $S$. The resulting DHs $H$
are found to be asymptotic to the event horizon in the distant
future. The next result is tailored to this scenario.  Let
$(\calM, g)$ be a dynamical black hole spacetime, as described in
section \ref{s4.2}. Recall, as defined there, a DH $H$ is said to
be asymptotic to the event horizon $E$ provided in the {\it
asymptotic region} (i.e., the region to the future of some
achronal spacelike hypersurface), the past Cauchy horizon $H^-(H)$
of $H$ coincides with $E$.

\begin{thm}\label{outermost2}  Suppose $H$  is a regular  DH
asymptotic to the event horizon $E$ in a spacetime $(\calM, g)$
obeying the NEC. Suppose, in the asymptotic region,  $H$ is
generated by a spacelike foliation $\{M_t\}$, such that each time
slice $M_t$ meets E in a single (connected) cross section.  Let
$S= S_{R_0}$ be the MTS of $H$ determined by the slice $M =
M_{t_0}$.  Then $S$ is the outermost WTS  in $M$.  That is, if
$\S$ is any other WTS in $M$, $\S$ does not meet the region of $M$
outside of $S$.
\end{thm}

\proof  By restricting to the asymptotic region, we may assume
that $H^-(H) = E$.  Let $\S$ be a MTS in $M$.  Since $M$ meets $E$
in a single cross section $S_E = M \cap E$, the portion of $M$
outside of $S_E$ is contained in the exterior region
$I^-(\scri^+)$.  Since  a WTS can not meet the exterior region,
$\S$ must be contained in the black region $B = \calM \setminus
I^-(\scri^+)$. If $\S$ touches $S_E$, which by the area theorem
has nonnegative expansion with respect to the outward null normal,
then the maximum principle Proposition~\ref{dhmaxprin} implies that
$\S = S_E$.   But this forces $S_E$ to have vanishing expansion,
contrary to the assumption we had made about dynamical black holes
(cf. section \ref{s4.2}).  Thus, $\S$ is actually contained in the
interior of the black hole region, $\S \subset {\rm int}\, B$.
Applying Theorem \ref{outermost} to the restricted foliation
$\{M'_t = M_t \cap {\rm int}\, B\}$, each leaf of which does not
meet $H^-(H) = E$, we conclude that $\S$ does not meet the region
of  $M$ outside of $S$, as was to be shown.\qed

In spacetimes with multiple black holes, the time slices $M_t$ may
meet the event horizon in several cross sections.  Theorem
\ref{outermost2} could still be applied to this multi-black hole
situation, by  suitably restricting each of the time slices $M_t$
to a subregion meeting the event horizon in a single cross
section.

Theorem \ref{outermost2}, together with several previous results, leads
to the following uniqueness result for dynamical horizons.

\begin{thm}\label{equal}  Suppose $H_1$ and $H_2$ are regular DHs
asymptotic to the event horizon $E$ in a black hole spacetime
$(\calM, g)$ obeying the NEC. Suppose, in the asymptotic region,
$H_1$ and $H_2$ are generated  by the same spacelike foliation
$\{M_t\}$, where each time slice $M_t$ meets E in a single
(connected) cross section. Then, in the asymptotic region, $H_1=
H_2$.
\end{thm}

\proof  \footnote{One may wish to argue as follows: By Theorem
\ref{outermost2}, in each time slice $M_t$, the MTSs of $H_1$ and
$H_2$ are both outermost, and hence agree.  However, implicit in this
argument is that the `outsides' determined by $H_1$ and $H_2$ are
compatible (e.g.  one cannot have the outside of an MTS of $H_1$
contained in the inside of an MTS of $H_2$.) For this and other
reasons we argue differently.} Without loss of generality we
may restrict spacetime $(\calM, g)$ to the asymptotic region.
We first show  that $H_1$ and $H_2$ meet. Since $H_1$ and
$H_2$ have a common past Cauchy horizon (namely E), it follows
that either $H_2$ meets $D^-(H_1)$ or $H_1$ meets $D^-(H_2)$.
Assume the latter holds.  By Corollary \ref{weave}, $H_1$ can not be
contained in $D^-(H_2) \setminus H_2$. Hence, $H_1$ must
meet $\partial D^-(H_2) = H_2 \cup H^-(H_2)$.  But $H^-(H_2) = H^-(H_1)$,
and $H_1$ can not meet its own past Cauchy horizon.  It follows that
$H_1$ meets  $H_2$.

Suppose $H_1$ and $H_2$ meet at $p \in M_{t_0}$.
Let $S_1$ and $S_2$ denote
the MTSs of $H_1$ and $H_2$, respectively, in the
time slice $M_{t_0}$.  Note that $p$ is on both $S_1$ and $S_2$.
Theorem~\ref{outermost2} applied to $H_1$ implies that
$S_2$ is in the region of $M_{t_0}$ inside of  $S_1$, yet meets
$S_1$ at $p$.
(Since $S_2$ is in the same component of  $M_{t_0} \cap B$ ($B = \calM \setminus I^-(\scri^+)$) as $S_1$, the added
condition on the time  slices $M_t$ in Theorem \ref{outermost2} is not needed
here; see
the remark after the proof of Theorem \ref{outermost2}.)
The maximum principle Proposition \ref{dhmaxprin}
then implies that  $S_2 = S_1$.  Now, by Corollary \ref{equal0},
$H_1 = H_2$.  \qed

\smallskip
In fact, Theorem \ref{equal} and its proof do not involve the event horizon
$E$ in an essential way.  It is sufficient to require that $H_1$ and $H_2$ be
asymptotic to each other, by which is meant that in the asymptotic region
(the region to the future of some achronal spacelike hypersurface)
the past Cauchy horizons of $H_1$ and $H_2$ agree.   In this more general
situation, one may apply  Theorem \ref{outermost} in lieu of Theorem \ref{outermost2}.

Recently, numerical simulations were performed specifically to
explore the properties of  MTTs \cite{num3} generated by a
space-like foliation. In the case of the gravitational collapse
leading to a single black hole, one begins with the foliation
$M_t$ in the asymptotic region and searches for MTSs on each leaf.
In some cases, each $M_t$ admits two well separated MTSs. Thus,
the MTTs obtained by evolution have two branches. The outer MTT is
a DH which is asymptotic to an isolated horizon while the inner
one is a timelike membrane (TLM). Theorem \ref{outermost2} ensures
that on any one time slice $M_t$, there is no MTS outside the
intersection $S_{R(t)}$ of $M_t$ and $H$. Our next result concerns
the TLM where the situation turns out to be just the opposite.

Let us first describe in somewhat more detail the setting for this
result.  Consider a TLM\,\, $T$ whose MTSs $S_t$ are determined by
a spacelike foliation $\{M_t\}$, i.e., $S_t = M_t \cap T$, such
that $S_t$ separates $M_t$ into an  `inside' (the side into which
the projection of $n$ points) and an `outside' (the side into
which the projection of $\ell$ points).  We will assume that the
time slices $M_t$ obey a very mild asymptotic flatness condition:
there exists outside each $S_t$ in $M_t$ a closed surface $\tilde
S_t$, such that it and $S_t$ form the boundary of a compact
$3$-manifold $V_t \subset M_t$, and such that $\tilde S_t$ has
positive expansion with respect to its outward null normal,
$\tilde \theta_t > 0$. Finally, we will assume that $T$ satisfies
the genericity condition that  the quantity
$\sigma_{ab}\sigma^{ab} + T_{ab}\ell^a \ell^b$ never vanish on
$T$, cf., section \ref{s2.2}. Raychaudhuri's equation, together
with the NEC, then implies that $\calL_{\ell} \tl < 0$ along each
MTS $S_t$.

\begin{prop} \label{tlm}  Let $T$ be a TLM in a spacetime $(\calM,g)$
satisfying the dominant energy condition.  Let $T$ be determined by a spacelike
foliation $\{M_t\}$, such that the time slices $M_t$ obey the
mild asymptotic flatness condition described above.  Assume further
that $T$ obeys the genericity condition.  Then there exists an
OMT surface outside of each
MTS $S_t$ in $M_t$.
\end{prop}

\proof The proposition follows from a recent result of Schoen
\cite{rs}. Very briefly, consider an MTS $S_{t'}$ slightly to the
past of the MTS $S_t$.  For $t'$ sufficiently close to $t$, the
null hypersurface generated by $\ell$ along $S_{t'}$, will meet
$V_t$ in a smooth closed surface $\hat S_t$ just outside $S_t$,
such that it and $\tilde S_t$ form the boundary of a compact
$3$-manifold $B_t \subset V_t$.  Moreover, taking $t'$ closer to
$t$ if necessary, the genericity condition will imply that $\hat
S_t$ has negative expansion, $\hat \theta_t < 0$.  Together with
the fact that $\tilde S_t$ has positive expansion $\tilde\theta_t
>0$, Schoen's result (which requires the dominant energy condition)
implies the existence of an OMT surface in the interior of $B_t$
(homologous, in fact, to $S_t$). \qed

Theorem \ref{tlm} says that the existence of the TLM $T$ by
itself implies that each $M_t$ must admit another OMT which lies
outside $M_t\cap T$. There is no a priori guarantee that the world
tube of these new OMTs would be a DH. But Theorem \ref{outermost}
suggests that the outermost MTT would be a DH.
Together, Proposition \ref{tlm} and Theorem \ref{outermost} imply
that the findings of numerical simulations \cite{num3} mentioned
above are not accidental but rooted in a general structure.

\section{Dynamical horizons and Killing vectors}
\label{s6}

We will now discuss some restrictions on the type of dynamical
horizons that can exist in spacetimes admitting isometries. We
will limit ourselves only to a few illustrative results that show
that some of the intuitive expectations on the interplay between
Killing symmetries and dynamical horizons are immediate
consequences of the main results of sections \ref{s3} and
\ref{s4}.

First, since DHs are associated with evolving black holes, one
would expect that stationary spacetimes do not admit DHs. This
conjecture is supported by a result which holds in the more
general setting of spacetimes admitting isolated horizons
$\Delta$. We will say that a DH $H$ is \emph{asymptotic to an
isolated horizon} $\Delta$ if there exists an achronal spacelike
hypersurface $V$ meeting $H$ in some MTS $S$ such that, in the
portion of spacetime to the future of $V$, the Cauchy horizon of
$H$ coincides with $\Delta$, i.e., $H^-(H) \cap I^+(V) = \Delta
\cap I^+(V)$. We then have:

\begin{thm}\label{nodh} Let $(\calM, g)$ be a spacetime obeying
the NEC which contains a non-expanding horizon $\Delta$. Suppose
$\Delta$ admits a cross-section $S$ such that the expansion of the
inward pointing null normal $n$ is negative and the expansion of
the outward pointing null normal $\ell$ satisfies $\mathcal{L}_{n}
\theta < 0$. Then $\calM$ does not admit any regular DH $H$
asymptotic to $\Delta$.
\end{thm}

\proof: Suppose $\calM$ does admit such a DH $H$. Then, by
displacing $S$ infinitesimally along $n$ we would obtain a WTS in
$D^-(H) \setminus H$ contradicting the result of Theorem \ref{basic1}. \qed

Event horizons of Kerr-Newman spacetimes are isolated horizons
satisfying the conditions of Theorem \ref{nodh}. Therefore, there
are no DHs in these spacetimes which asymptote to their event
horizons. More generally, on physical grounds one expects the
conditions to be met on event horizons $E$ of any stationary black
hole, including those in which $E$ may be distorted due to
presence of external matter rings or non-Abelian hair. This
expectation is borne out in a large class of distorted static,
axi-symmetric black holes solutions which are known analytically
(see, e.g., \cite{afk}). Next, note that Theorem \ref{nodh} does
not require the presence of a stationary Killing field. Therefore,
it rules out the existence of DHs of the specified type also in
certain radiating spacetimes. An explicit example of this
situation is provided by a class of Robinson-Trautman spacetimes
which are vacuum, radiating solutions which nonetheless admit an
isolated horizon \cite{pc}. Finally, note that the precise sense
in which the DH is assumed to be asymptotic to an IH is important.
Some Vaidya metrics, for example, admit DHs $H$ which are smoothly
joined onto IHs $\Delta$ in the future \cite{ak3}. However, in
these cases, $\Delta$ does not overlap with the past Cauchy
horizon $H^-(H)$, whence $H$ is not asymptotic to $\Delta$ in the
sense spelled out above.

In globally stationary black hole spacetimes, on the other hand,
Theorem \ref{nodh} does not fully capture the physical expectation
because it does not rule out the existence of `transient' DHs,
i.e, DHs which are not asymptote to event horizons. We will now
establish a few results which restrict this possibility. (See
also \cite{ms}.)

\begin{prop} \label{prop1} Let $(\calM, g)$ be a spacetime
satisfying the NEC. Let it admit a timelike or null Killing field
which is nowhere vanishing in some open region $R$. Then $R$ can
not contain a regular DH $H$. \end{prop}

This is a consequence of the following more general result.
\begin{prop} \label{prop2} Let $(\calM, g)$ be a spacetime satisfying
the NEC which admits a Killing field $K$ in some open region $R$.
Then, $R$ can not contain a regular DH $H$ such that $K$ is
everywhere transversal to $H$ on any one of its leaves (i.e.,
MTSs) $S$. \end{prop}

\proof: Suppose $K$ is transversal to $H$ on a leaf $S$. Then, the
image of $S$ under an infinitesimal isometry generated by $K$
(with appropriate orientation) would provide an MTS in
$D^-(H)\setminus H$. This would contradict Theorem \ref{basic1}.
\qed

Finally, let us consider the complementary case in which the
Killing field is tangential to $H$. There is an interesting
constraint also in this case.

\begin{prop} \label{prop3} Let a Killing field $K$ be tangential to
a DH $H$. Then, $K$ is tangential to each leaf (i.e., MTS) $S$ of
$H$.\end{prop}

\proof Suppose there is a leaf $S$ to which $K$ is not everywhere
tangential. Then, the images of $S$ under isometries generated by $K$
would provide a 1-parameter family of MTSs $S_t$. By Corollary
\ref{cor1}, it follows that each $S_t$ must belong to the unique
MTS-foliation of $H$. However, because the expansion of $\ell$
vanishes and that of $n$ is everywhere negative, the area of
leaves of this foliation changes monotonically. On the other hand,
since each $S_t$ is the image of $S$ under an isometry, its area
must be the same as that of $S$. This contradiction implies that
our initial assumption can not hold and $K$ must be tangential to
each leaf of the foliation.\qed

While these results serve to illustrate the interplay between DHs
and Killing vectors, it should be possible to strengthen them
considerably. Physically, the most interesting case is when the
topology of each leaf $S$ of $H$ is that of a 2-sphere and $K$ is
spacelike. Then, if $K$  is tangential to $H$, Proposition
\ref{prop3} ensures that it must be a rotation. Proposition
\ref{prop2} implies that $K$ can not be everywhere transversal to
$H$ even on a single leaf $S$. What is missing is an exhaustive
analysis of the intermediate case where $K$ is partly tangential
and partly transverse to $H$.

\section{Discussion}
\label{s7}

As explained in sections \ref{s1} and \ref{s5}, in numerical
relativity one starts with initial data, makes a judicious choice
of a time function (more precisely, of lapse and shift fields) and
evolves the data using Einstein's equations. The resulting
space-time $(\calM,g)$ is then naturally equipped with a foliation
$M_t$. In simulations of black hole space-times, one finds
marginally trapped tubes (MTTs) some of which are DHs. Since the
whole construction is tied to a foliation $M_t$, one might first
think that by varying the foliation, one would be able to generate
an uncontrollably large class of DHs even in the portion of
space-time containing a single black hole. However, there exist
heuristic arguments to suggest that a vast majority of these
foliations would lead to timelike membranes rather than DHs
\cite{de}, whence the number of DHs would be restricted. By
establishing precise constraints on the occurrence and location of
DHs, we have shown that this expectation is borne out to a certain
extent. In particular, the DHs are not so numerous as to provide a
foliation of a space-time region even locally; given a DH $H$,
there is no other DH to its past in $D(H)$; and, the DHs of
numerical simulations can not bifurcate. Our uniqueness results
also imply that DHs which are asymptotic to the event horizon are
non-trivially constrained, providing an `explanation' of the fact
that numerical simulations invariably lead to an unambiguous DH
with this property \cite{num2,num3,bgvdb,acg}.

However, considerable freedom still remains; our uniqueness
results are quite far from singling out a canonical DH in each
connected, inextendible trapped region ${\mathcal{T}}$ of
space-time, representing a single evolving black hole. Can these
results be strengthened, perhaps by adding physically reasonable
conditions? One such strategy was suggested by Hayward
\cite{sh1,sh2}. Using seemingly natural but technically strong
conditions, he argued that the dynamical portion of the boundary
$\partial {\mathcal{T}}$ of ${\mathcal{T}}$ would be a DH $H$.
This $H$ could serve as the \emph{canonical} DH associated with
that black hole. However, Hayward's assumptions may be too strong
to be useful in practice. To illustrate the concerns, let us
consider an asymptotically flat space-time containing an event
horizon $E$ corresponding to a single black hole. There is a
general expectation in the community that, given any point in the
interior of $E$, there passes a (marginally) trapped surface
through it (see, e.g., \cite{de}). This would imply that the
boundary $\partial {\mathcal{T}}$ of the trapped region is the
\emph{event horizon} which, being null, can not qualify as a
dynamical horizon.

In the general context, arguments in favor of this expectation
have remained heuristic. However, in specific cases, such as the
Vaidya metric or spherical collapse of scalar fields, it should be
possible to make progress. In these cases, $H$ is well separated
from $E$. As one would expect from Hayward's scenario, no
\emph{round} trapped surface meets the region outside $H$.
However,  certain arguments, coupled with a recent result of
Schoen \cite{rs}, suggest that non-round trapped surfaces can
enter this region and the boundary of the trapped region is
therefore the event horizon. (More precisely, these arguments
imply that an outer marginally trapped surface enters the outside
of $H$.) If this suggestion is borne out, one would conclude that
Hayward's strategy to single out a canonical DH is not viable.
However, this would not rule out the possibility of singling out a
canonical dynamical horizon through some other conditions. Indeed,
the Vaidya solutions do admit a canonical DH but so far it can be
singled out only by using the underlying spherical symmetry. Is
there another, more generally applicable criterion?

The DH framework also provides an avenue to make progress on the
proof of the  Penrose conjecture beyond the context of
time-symmetric initial data. Consider an asymptotically flat
space-time $(\calM, g)$ the distant future region $R$ of which
contains a single black hole. Let us suppose that there is a DH
$H$ which is generated by a foliation $M_t$ in the asymptotic
region. Then, by Theorem \ref{outermost2}, each MTS $S_t$ of $H$
is the outermost MTS in $M_t$. Now, the area radius $R_t$ of $S_t$
increases monotonically whence $R_t < R_{i^+}$, where $R_{i^+}$ is
the asymptotic value of the area radius of MTSs of $H$ \cite{ak2}.
The DH framework assigns a mass to each MTS whose expression
implies $R_{i^+} \le 2GM_{i^+}$ (where $G$ is Newton's constant)
\cite{abl2,ak3}.  Finally, when $H$ is asymptotic to $E$, on
physical grounds one expects $M_{i^+}$ to equal the future limit
$M^{\rm Bondi}_{i^+}$. This would imply a generalized Penrose
inequality $R_t < 2GM^{\rm Bondi}_{i^+}$ relating the future limit
of the Bondi mass on $\scri^+$ to the area of the outermost
marginally trapped surface on any leaf $M_t$ of the foliation.
Thus, the missing step in the proof is to establish the physically
expected relation between $M_{i^+}$ and the future limit $M^{\rm
Bondi}_{i^+}$ of the Bondi mass. The resulting theorem would have
to make two sets of assumptions: the existence of a DH $H$ which
asymptotes to the event horizon $E$ in an appropriate sense and
decay properties of gravitational radiation and matter fluxes
along $E$ and $\scri^+$ in the asymptotic future. These are rather
different from the assumptions that go into the well-established
partial results on Penrose inequalities. However, they are
physically well-motivated.  Furthermore, a result along these
lines would not have to ask that the initial data on $M_t$ be
time-symmetric, and the result would be stronger than those
discussed in the literature in that it would refer to the future
limit $M^{\rm Bondi}_{i^+}$ of the Bondi mass (which is strictly
less than the Arnowitt-Deser-Misner mass in dynamical situations).

\section*{Acknowledgments}

We thank Lars Andersson, Robert Bartnik, Jim Isenberg, Badri
Krishnan, Eric Schnetter, and Chris Van Den Broeck for discussions
and correspondence and Parampreet Singh and Tomasz Pawlowski for
help with figures. This work was supported in part by NSF grants
DMS-0104042, PHY-0090091, and PHY-0354932, the Alexander von
Humboldt Foundation, the C.V. Raman Chair of the Indian Academy of
Sciences and the Eberly research funds of Penn State.

\bibliographystyle{my-h-elsevier}

\end{document}